\documentclass[prd, aps, superscriptaddress, preprintnumbers, twocolumn, floatfix, nofootinbib]{revtex4-2} 
\pdfoutput=1

\usepackage{amsfonts}
\usepackage{amsmath}
\usepackage{amssymb}
\usepackage{bm}
\usepackage{dcolumn}
\usepackage{graphicx}
\usepackage[latin1]{inputenc}
\usepackage{latexsym}
\usepackage{rotating}
\usepackage{hyperref}
\usepackage{graphicx}
\usepackage{color}

\newcommand\be{\begin{equation}}
\newcommand\bea{\begin{eqnarray}}
\newcommand\ee{\end{equation}}
\newcommand\eea{\end{eqnarray}}

\renewcommand{\d}{{\mathrm{d}}}
\renewcommand{\[}{\left[}
\renewcommand{\]}{\right]}
\renewcommand{\(}{\left(}
\renewcommand{\)}{\right)}
\newcommand{\nn}{\nonumber}

\def\doi{http://doi.org}

\def\d{\mathrm{d}}

\begin{document}

\title{Emergent Cosmology from Matrix Theory}

\author{Suddhasattwa Brahma}
\email{suddhasattwa.brahma@gmail.com}
\affiliation{Department of Physics, McGill University, Montr\'{e}al, QC, H3A 2T8, Canada}

\author{Robert Brandenberger}
\email{rhb@physics.mcgill.ca}
\affiliation{Department of Physics, McGill University, Montr\'{e}al, QC, H3A 2T8, Canada}

\author{Samuel Laliberte}
\email{samuel.laliberte@mail.mcgill.ca}
\affiliation{Department of Physics, McGill University, Montr\'{e}al, QC, H3A 2T8, Canada}

\date{\today}

\begin{abstract}
 
Matrix theory is a proposed non-perturbative definition of superstring theory in which space is emergent. We begin a study of cosmology in the context of matrix theory. Specifically, we show that matrix theory can lead to an emergent non-singular cosmology which, at late times, can be described by an expanding phase of Standard Big Bang cosmology. The horizon problem of Standard Big Bang cosmology is automatically solved. We show that thermal fluctuations in the emergent phase source an approximately scale-invariant spectrum of cosmological perturbations and a scale-invariant spectrum of gravitational waves. Hence, it appears that matrix theory can lead to a successful scenario for the origin of perturbations responsible for the currently observed structure in the universe while providing a consistent UV-complete description.
 
\end{abstract}

\maketitle

\section{Introduction}

The inflationary scenario \cite{Guth} is not the only early universe paradigm consistent with current observations. As already pointed out a decade before the development of the inflationary scenario \cite{early}, what is required in order to explain the origin of acoustic oscillations in the angular power spectrum of the cosmic microwave background (CMB) and the {\it Baryon Acoustic Oscillations} in the matter power spectrum is an early phase in the evolution of the universe which generates an approximately scale-invariant spectrum of nearly adiabatic and nearly Gaussian curvature fluctuations. Inflation is one way to obtain such a spectrum \cite{Mukh}, but there are others (see e.g. \cite{RHBrev} for a comparative review of several scenarios). 

Specifically, in \cite{BV} an {\it emergent scenario} was proposed in which the universe originates in a quasi-static phase (``Hagedorn phase'') of a gas of strings at a temperature close to the Hagedorn temperature \cite{Hagedorn}, the limiting temperature of a gas of closed strings \footnote{In the following, we will call this scenario {\it String Gas Cosmology}.}. Via a phase transition (which in String Gas Cosmology is determined by the decay of string winding modes, or more generally speaking, by the spontaneous breaking of the T-dual symmetry of the state), this emergent phase connects to the radiation phase of Standard Big Bang (SBB) cosmology. As shown in \cite{NBV, NBPV} (see \cite{RHBSGrev} for a review), thermal fluctuations of the gas of strings lead to an almost scale-invariant spectrum of cosmological fluctuations with a slight red tilt \cite{NBV}, and of gravitational waves with a slight blue tilt \cite{NBPV}. Additionally, the String Gas scenario yields a non-singular cosmology. Note that, unlike in inflationary cosmology where the cosmological fluctuations (which are observed today) are generated in the early universe in their quantum vacuum state, in String Gas Cosmology the initial state is assumed to be a thermal one and the fluctuations are hence of thermal origin \footnote{Note that there is a variant of standard inflation, namely {\it warm inflation} \cite{warm} in which the fluctuations also emerge thermally.}. However, in \cite{BV} no dynamics for the Hagedorn phase were provided. 

Recently, a very interesting proposal appeared \cite{Vafa} postulating that the early phase is a topological phase, and demonstrating that the predictions of String Gas Cosmology for the spectrum of curvature fluctuations can be recovered. Here, we propose an alternative view of the emergent phase, a model based on {\it matrix theory}. In analogy to what is assumed in String Gas Cosmology, the fluctuations are of thermal origin, and lead to power spectra of scalar and tensor modes which are consistent with current observations.

Matrix theory is the suggestion that  certain large $N$ matrix models can provide non-perturbative definitions of superstring theory. There are two main proposals for matrix theory, the BFSS model \cite{BFSS} and the IKKT proposal \cite{IKKT} (see \cite{Ydri} for a recent review of these and other matrix models). In the BFSS model, the matrices are functions of time, and space emerges from the properties of the matrices which will be discussed in the following section \footnote{See also the $c = 1$ matrix model of \cite{Sumit} which yields a non-critical string theory with one spatial dimension.}. In the IKKT proposal, time is emergent as well. The scenarios are related in the sense that the high temperature limit of the BFSS model yields the IKKT action (compactification on a thermal circle).

Our starting point will be the BFSS model, conjectured to be the non-perturbative proposal for M-theory. Making use of the equivalence between the high temperature limit of the BFSS model and the IKKT action, and starting in a thermal state, we will use the results of detailed numerical studies of the IKKT model \cite{IKKTE} to show that a background develops in which there is a separation between three spatial dimensions which become large, and six which remain compact, similar to what was argued to happen in String Gas Cosmology \cite{BV}. In this background, we then compute the thermal fluctuations of the energy-momentum tensor and determine the resulting spectra of curvature fluctuations and of gravitational waves.
 
We will consider a $D = d + 1$ dimensional space-time, where $d$ is the number of spatial dimensions (which is $d = 9$ for superstring theory and $d=3$ after the phase transition in the IKKT model). Roman indices will be used to refer to the spatial directions. When discussing late time cosmology, we denote the cosmological scale factor by $a(t)$, where $t$ is physical time, and use comoving spatial coordinates $x$. The Hubble expansion rate is given by $H \equiv {\dot{a}}/a$, an overdot denoting the derivative with respect to time. The inverse of $H$ is the Hubble radius which plays a key role in the evolution of cosmological perturbations. As usual, the commutator of two matrices $A$ and $B$ is denoted by $[A, B]$.

\section{Background}

Our starting point will be the BFSS matrix model \cite{BFSS}. The basic objects in this model  are $d = 9$ bosonic $N \times N$ Hermitian matrices $X^i_{a, b}$ and their sixteen fermionic superpartners $\theta_{a, b}$ which transform as spinors under the ${\rm SO}(9)$ group of spatial rotations. There is a $U(N)$ gauge symmetry, and $A$ is the associated gauge field, another $N \times N$ matrix. The Lagrangian is given by
\bea \label{BFSSaction}
L \, &=& \, \frac{1}{2 g^2} \bigl[ {\rm Tr} \left \{\frac{1}{2} \(D_t X_i\)^2 - \frac{1}{4} \[X_i, X_j\]^2\right\} \nonumber \\
&& - \theta^T D_t \theta - \theta^T \gamma_i \[ \theta, X^i \] \bigr] \, 
\eea
where $D_t :=\partial_t - i\[A(t), \cdot\]$ is the usual covariant derivative.  The large-$N$ limit corresponds to taking $N \rightarrow \infty$ while holding the 't Hooft coupling $\lambda := g^2 N$ fixed. The proposal of \cite{BFSS} is that in the large-$N$ limit the action (\ref{BFSSaction}) yields a non-perturbation definition of M-theory (see \cite{Taylor_review} for a detailed review of this proposal using the discrete light-cone quantization in the `infinite-momentum frame'). In particular, in this limit space is emergent. The spatial coordinates are related to the eigenvalue distribution of the matrices $X^i$ in a similar way to how the one spatial dimension arises in the $c = 1$ matrix model of non-critical string theory \cite{Sumit}.

As mentioned, there are $d+1$ bosonic matrices $A(t),\,X_i(t)\,\; (i=1,2,\ldots,d)$, each of which is an $N \times N$ Hermitian matrix. At finite temperature $T$, the bosonic part of the BFSS action is given by
\be \label{BFSSbosonic}
S(\beta) \, = \, \frac{1}{g^2} \int_0^{\beta} \d t\; {\rm Tr}\left\{\frac{1}{2} \(D_t X_i\)^2 - \frac{1}{4} \[X_i, X_j\]^2\right\}\,,
\ee
where $\beta = 1/T$.  One can set $\lambda=1$ without any loss of generality, and thus we can trade $1/g^2$ for $N$ in front of the action. We choose a unit convention in which the mass dimension of $X$ is $1$ and that of $g^2$ is $3$ (we have set $l_s =1$ for now).

At high temperatures, the BFSS model reduces to the IKKT model \cite{equiv, equiv3}. Since we will use results from studies of the IKKT setup to establish our cosmological background, we recall the key points of the latter model.

The IKKT matrix model \cite{IKKT, IKKTE} (see \cite{Nishimura} for a recent review) is defined by the following action
\be \label{IKKTaction}
	S \, = \,  -\frac{1}{g^2} \text{Tr}\(\frac{1}{4} \[A^a, A^b\]\[A_a,A_b\] + \frac{i}{2} \bar{\psi}_\alpha \({\cal{C}} \Gamma^a\)_{\alpha\beta} \[A_a,\psi_\beta\]\)\,,
\ee
where $\psi_\alpha$ and $A_a$ ($\alpha =1,\ldots,16$, $a=1,\ldots,10$) are $N\times N$ fermionic and bosonic Hermitian matrices, respectively, the $\Gamma^{\alpha}$ are the gamma-matrices for $D = 10$ dimensions, and ${\cal{C}}$ is the charge conjugation matrix. While $a$ is a ten-dimensional vector index, $\alpha$ is a spinor index such that $\psi_\alpha$ plays the role of a ten-dimensional Majorana-Weyl spinor. This action can be seen as a matrix regularization of the worldsheet action of Type IIB superstring theory in the Schild gauge. Depending on the metric which is used to raise and lower the indices (either Euclidean or Minkowski), the action can be viewed as that of the Euclidean or Lorentzian type IIB matrix model.  If we choose to have the same coupling $g^2$ for the IKKT model, as before, with mass dimension $3$, the matrices $A^a$ will have mass dimension $3/4$.

The action of the Euclidean matrix model \cite{IKKT} is given by the following functional integral over the bosonic and fermionic fields
\be
Z \, = \, \int dA d\psi e^{- S} \, ,
\ee
while the action of the Lorentzian model is defined by \cite{IKKTE}
\be
Z \, = \, \int dA d\psi e^{iS} \, .
\ee

In the IKKT approach, both space and time are emergent, time being related to the matrix $A_D$, while space results from the other matrices. The $SU(N)$ symmetry can be used to diagonalize the matrix $A_{10}$
\be
A_{10} \, = \, {\rm{diag}} (\alpha_1, ... , \alpha_N) \, ,
\ee
where without loss of generality the $\alpha_i$ can be ordered in ascending magnitude. As was shown in \cite{IKKTE}, the spatial matrices $A_i$ then have a band-diagonal structure in the sense that there is an integer $n \ll N$ such that the matrix elements $(A_i)_{ab}$ for $|a - b| > n$ are much smaller than those for $|a - b| \leq n$.  Note that $n$ is a fixed fraction of $N$ and hence goes to infinity in the limit $N \rightarrow \infty$.

A time variable $t$ (which will be the time variable of our emergent phase) can then be defined by averaging the diagonal elements $\alpha_i$ over $n$ elements
\be
t(m) \, \equiv \, \frac{1}{n} \sum_{l = 1}^n \alpha_{m + l} \, ,
\ee
where the index $m$ runs from $1$ to $N - n$. Time-dependent spatial matrices $({\bar{A_i}})_{I, J}(t)$ of dimension $n \times n$ can then be defined via
\be
({\bar{A_i}})_{I, J}(t(m)) \, \equiv \, (A_i)_{m + I, m + J} \, .
\ee
In the limit $N \rightarrow \infty$ we have $n \rightarrow \infty$ and $({\bar{A_i}})_{I, J}(t(m))$ becomes the emergent continuum space.

It is then natural to define the extent $x_i$ of a given spatial dimension $i$ at time $t$ by
\be
x_i(t)^2 \, \equiv \, \left\langle \frac{1}{n} \text{Tr} ({\bar{A_i}})(t))^2 \right\rangle \, ,
\ee
where the pointed brackets stand for the quantum expectation value in the state defined by the partition function. Then, the total extent $R(t)$ of space at time $t$ is 
\be\label{IKKT_R}
R^2(t) \, = \, \sum_{i = 1}^9 x_i(t)^2 \, .
\ee
Following \cite{Nishimura}, it is more convenient to define the {\it moment of inertia tensor}
\be
T_{ij} \, \equiv \, \left\langle \frac{1}{n} \text{Tr} ({\bar{A_i}})(t) {\bar{A_j}})(t)) \right\rangle  
\ee
which is a symmetric $9 \times 9$ matrix whose eigenvalues can be denoted by $\lambda_i$.

A numerical analysis of this system shows \cite{IKKTE} that as a function of time the $SO(9)$ spatial symmetry is spontaneously broken. Of the nine eigenvalues $\lambda_i(t)$, three of them become large, while six remain close to the original size. This is the same symmetry breaking pattern obtained in String Gas Cosmology \cite{BV} from considerations of the annihilation of string winding modes which can only liberate three spatial dimensions. What is observed in matrix theory can be viewed as the non-perturbative picture of the scenario of \cite{BV}.

A similar symmetry breaking pattern was first studied in the Euclidean framework. There, the emergence of three large spatial dimensions can be seen both numerically and using a Gaussian expansion method in which the free energy is computed when approximating the functional integral via a Gaussian expansion about particular configurations, and the resulting free energy is compared for different chosen configurations, to find that the free energy is minimized for $d=3$. See \cite{IKKT2} for a selection of papers on this topic. On the other hand, the analyses for the Lorentzian IKKT model is more subtle. Firstly, numerical investigations are much more technically involved due to the `sign problem'. Although this was averted using some approximations involving assuming a Gaussian action for the bosonic part of the IKKT action in \cite{IKKTE}, it was soon realized that the expanding spacetime has only two independent large eigenvalues (the so-called Pauli-matrix structure) \cite{Pauli_structure}. In subsequent work, this was found to be a pathology of the approximation which was used to solve the sign problem and the numerical `complex Langevin method' was introduced using two parameters to denote Wick rotations on both the worldsheet and the target space. This culminated in finding a true (smooth) $(3+1)$-d emergence of the background from the Lorentzian IKKT model \cite{CLM}, where the presence of fermionic matrices turn out to be essential.

In the high temperature limit, the BFSS model reduces to the (Euclidean) IKKT scenario. The BFSS gauge field matrix $A(t)$ corresponds to the IKKT matrix $A_D(t)$ and the BFFS spatial matrices $X_i(t)$ become the matrices $A_i(t)$ in the IKKT model. Hence, taking the results from the analysis of the IKKT model described above back to the BFSS side, we argue that in the high temperature limit (which is relevant for our discussion of the emergent phase) the background which minimizes the free energy will experience spontaneous symmetry breaking in which three of the spatial dimensions (given by the quantum expectation values of eigenvalue distribution of the matrices $X_i$) become large compared to the other six. Note that this symmetry breaking occurs during the emergent phase, and not only at the end of it. 

Irrespective of whether one begins with the Euclidean or the Lorentzian version of the IKKT model, this emergence can be understood more generally.  Since the eigenvalues of the matrices denote the target space coordinates, at very early times before the symmetry-breaking phase transition takes place, the nine eigenvalues are of equal size and of a microscopic scale, on which there does not exist a smooth geometric picture of spacetime. One way to see this is that the eigenvalues denote positions of D-branes and the matrices, corresponding to these, commute only in the limit after the symmetry breaking when the eigenvalues of three matrices become large. The target space coordinates are inherently non-commutative at very early times. This is a \textit{non-geometric phase} from which our $(3+1)$-d universe emerges in the matrix model. As the emergent phase proceeds and the three large spatial dimensions grow in size, we will reach a point when the effective field theory description via Einstein gravity yields a good approximation for the infrared modes which we are interested in when considering cosmological perturbations measured at late times. Let us for now consider this transition to take place at a fixed time $t_c$, and we return to a discussion of how this transition happens later on. 

Let us summarize the background cosmology which we are proposing. We start in a high temperature state of the BFSS matrix model which is equivalent to the IKKT matrix model (as $T\rightarrow \infty$). After Wick rotating the IKKT model, we can diagonalize the $A_{10}$ matrix, and the diagonal elements determine our emergent time variable. The diagonal blocks of the $A_i, i = 1, .., 9$ matrices in this basis define an evolving space. All spatial dimensions (measured as described above in terms of the expectation values of the spatial matrices) are of the typical microscopic scale (the string scale). The spatial matrices evolve in time and the emergent space undergoes symmetry breaking in which three spatial dimensions become large and the other ones remain microscopic. The $SO(9)$ symmetry of space is broken to $SO(3) \times SO(6)$. As the three-dimensional space expands, General Relativity becomes a good description of the low energy dynamics of the three dimensional space \footnote{Note that the Lorentz symmetry of the effective theory is a result of the $SO(10)$ symmetry of the original Euclidean IKKT matrix model.} A transition to the expanding phase of Standard Big Bang cosmology occurs. Our subsequent analysis of fluctuations is independent of the specifics of this transition in the same way that the analysis of fluctuations in inflationary cosmology are in general insensitive to the details of reheating.
 
At this point, let us give some justification for using General Relativity (GR) as the low-energy limit of the matrix theory. How do we know that the low-energy gravity theory is going to be GR and not something else? Firstly, note that the BFSS model is a proposal for M-theory and therefore, we are guaranteed to have GR as the low-energy limit of (the gravitatiuonal part of) this theory. Furthermore, within the IKKT model, it has been shown that the underlying diffeomorphism symmetry of GR emerges naturally from this \cite{IKKT_effective_Spacetime}. But the most direct way to note this was shown within the operator interpretation of matrices in the IKKT model, in which one could derive the vacuum Einstein equations starting from the classical equations of motion of the IKKT model \cite{IKKT_operator}. In this approach, matter and gauge fields appear as fluctuations on top of a gravitational background and thus all fields of different spins, including the graviton, emerge from the same IKKT model. Going to higher-order corrections, one can find different quantum fields sourcing Einstein's equations. However, the exact dynamics of quantum fields are yet to be understood in this interpretation and, therefore, we approach the problem from a different perspective. We consider a natural state for our cosmological model, namely a thermal state, which yields the emergent background from the IKKT model discussed above. Since the state is a thermal state, it includes thermal fluctuations which yield source terms for late time cosmological perturbations. Based on the above arguments, and the fact that after the time $t_c$ we are in the low-energy limit of superstring theory, we use Einstein's equations sourced by the thermal state to compute the cosmological perturbations whose properties we calculate below.

In contrast to Standard Cosmology and the Inflationary paradigm, our proposed cosmology does not suffer from an initial singularity problem because the early phase is a nonsingular quantum mechanical matrix model which cannot be described by Einstein gravity. In particular, there is no beginning of time in the sense of General Relativity. Furthermore, recall that the BFSS matrix model is a quantum mechanical model and does not suffer from field-theoretic divergences one has to contend within GR.

The origin of our proposed cosmology as a quantum mechanical matrix model also provides a solution to the Horizon Problem of Standard Big Bang cosmology. The initial thermal state of the quantum mechanical matrix model automatically generates correlations over the entire emergent spatial section. From the point of view of an emergent scenario it is very reasonable to assume that we start in a thermal state \footnote{Note that it is only in  the high temperature limit that the correspondence between the BFFS and IKKT models has been established.}. Thus, like in String Gas Cosmology, the cosmological fluctuations and primordial gravitational waves will be of thermal origin, unlike in inflationary cosmology where the inhomogeneities emerge from quantum vacuum fluctuations. 

Let us also point out a difference between our cosmological model emerging from matrix theory and String Gas Cosmology. Unlike the latter, our model cannot be thought of as a \textit{free} collection of stringy objects, such as $D$-branes, whose thermal properties describe the thermal state sourcing cosmological perturbations. It is tempting to interpret our results as a collection of `free' $D0$-branes in a box since our starting point is the BFSS model. Similarly, one might be led to presume that a box of `free' $D(-1)$ branes would explain the background dynamics due to its origins in the IKKT model. If this were to be true, one could have studied the thermodynamics of free $D$-branes just like one does for a box of strings in String Gas Cosmology. However, the crucial point to realize is that the BFSS (or, similarly, the IKKT) model is not just any collection of $D0$-branes but a specific bound state configuration of them. This gives rise to a very specific theory which allows us to do our computations in a thermal state that, quite remarkably, gives rise to scale-invariant perturbations in the early-universe, as we shall show later on. In fact, the thermal properties of a collection of free $D0$-branes do not have the same properties as can be seen from \cite{D0_branes_thermodynamics}.
 
The cosmological fluctuations and gravitational waves which we can measure today in cosmological experiments have a length scale which even at the end of the emergent phase is in the far infrared compared to the typical energy scale of the emergent phase. Hence, the evolution of fluctuations on these scales will be described by the usual linear cosmological perturbation theory based on Einstein gravity. Hence, as in the case of String Gas Cosmology \cite{RHBSGrev}, the metric fluctuations will be determined by the correlation functions of the energy-momentum tensor in the thermal state of the emergent phase. In the following section we turn to the computation of these fluctuations.
 
\section{Fluctuations}

\subsection{Formalism}

In the previous section we have described our model for the background of the emergent period which results from matrix theory. Via a phase transition, the emergent period will connect to the radiation phase of the SBB in $3 + 1$ space-time dimensions. In this section we will compute the spectra of cosmological fluctuations and gravitational waves which arise from our background. In the framework of an emergent cosmology, and in contrast to the situation in an inflationary model, the length scales on which we currently observe the fluctuations were always many orders of magnitude larger than the typical microscopic scales, e.g. the Planck length. Specifically, if the energy scale at which the transition from the emergent phase to the radiation phase of the SBB occurs is $10^{16} {\rm GeV}$, then the wavelengths at that time were of the order of $1 {\rm mm}$ or larger. Hence, the description of fluctuations using the usual theory of linear cosmological perturbations (see e.g. \cite{MFB, RHBpertRev} for reviews) will apply.

We will write the metric of our four dimensional space-time (which we assume to be spatially flat) in longitudinal gauge, i.e. in the form
\be
ds^2 \, = \, (1 + 2 \Phi) dt^2  - a(t)^2 \[ (1 - 2 \Phi) \delta_{i j} + h_{ij} \] dx^i dx^j \, ,
\ee 
where $\Phi(x, t)$ is the relativistic generalization of the Newtonian gravitational potential, and the transverse and traceless tensor $h_{i j}$ represents the gravitational waves. Specifically, a gravitational wave with dimensionless polarization tensor $\epsilon_{i j}$ will have an amplitude $h(x, t)$. We are neglecting the contribution of vector modes since these modes decay in an expanding universe.

Note that in a thermal state, the fluctuations on the typical microscopic state may be large in amplitude, but on the infrared scales relevant to cosmological observations they will be Poisson suppressed and hence small in amplitude such that linear cosmological perturbation theory applies and all Fourier mode of the fluctuating fields evolve independently.

According to the theory of linear cosmological perturbations, the curvature fluctuation on a scale $k$ (where $k$ denotes comoving wave number) is given by the energy density perturbations on that scale via
\be \label{scalarfluct}
\langle\vert \Phi(k) \vert^2\rangle \, = \, 16 \pi^2 G^2 k^{-4} \langle \delta T^0_0(k) \delta T^0_0(k) \rangle \, ,
\ee
where $T^{\mu}_{\nu}$ is the energy-momentum tensor of matter (also evaluated in longitudinal gauge), and $G$ is Newton's gravitational constant. Similarly, the amplitude $h(k)$ of a gravitational wave mode is determined by the off-diagonal pressure fluctuations via \footnote{The notation is a bit loose here: the indices $i$ and $j$ correspond to the polarization state of the gravitational wave.}
\be \label{tensorfluct}
\langle\vert h(k) \vert^2\rangle \, = \, 16 \pi^2 G^2 k^{-4} \langle \delta T^i_j(k) \delta T^i_j(k) \rangle \,,\,\, i \neq j
\ee

On sub-Hubble scales, matter fluctuations dominate over the induced curvature fluctuations. Hence, following the logic used in String Gas Cosmology in \cite{NBV, NBPV}, we will first use the partition function $Z$ of our model to determine the correlation functions of the energy-momentum tensor. Using these results, we apply (\ref{scalarfluct}) and (\ref{tensorfluct}) to determine the initial amplitude of the curvature fluctuations and gravitational waves when the length mode $k$ exits the Hubble radius at the end of the emergent phase. From then on until the present time the usual evolution of the cosmological fluctuations applies.

In a thermal state, the fluctuations in the energy-momentum tensor in a box of radius $R$ are determined in terms of the finite temperature partition function of the system. Specifically, since
\be
\langle T^{\mu}_{\nu}\rangle \, = \, 2 \frac{g^{\mu \lambda}}{\sqrt{-g}} \frac{\partial {\rm ln} Z}{\partial g^{\nu \lambda}} \, 
\ee
the fluctuations of the energy-momentum tensor in a box of radius $R$ are given by (see \cite{RHBSGrev} for details)
\bea
{C^{\mu}_{\nu}}^{\sigma}_{\lambda} \, &\equiv& \, \bigl< T^{\mu}_{\nu} T^{\sigma}_{\lambda} \bigr> - \bigl< T^{\mu}_{\nu} \bigr> \bigl< T^{\sigma}_{\lambda} \bigr> \\
&=& \, 2 \frac{g^{\mu \alpha}}{\sqrt{-g}} \frac{\partial}{\partial g^{\alpha \nu}} 
\bigr( \frac{g^{\sigma \delta}}{\sqrt{-g}} \frac{\partial {\rm ln}Z}{\partial g^{\delta \lambda}} \bigl) \nonumber \\
&& + 2 \frac{g^{\sigma \alpha}}{\sqrt{-g}} \frac{\partial}{\partial g^{\alpha \lambda}} 
\bigr( \frac{g^{\mu \delta}}{\sqrt{-g}} \frac{\partial {\rm ln}Z}{\partial g^{\delta \nu}} \bigl) \, , \nonumber
\eea
where $Z$ is the partition function restricted to the box. Specifically, the energy density fluctuations are determined by
\be \label{eflucts}
{C^{00}}_{00} \, = \delta{\rho}^2 \, = \frac{T^2}{R^6} C_V
\ee
where $C_V$ is the specific heat capacity in a box of radius $R$ and is given by the partial derivative of the internal energy $E(\beta)$ with respect to temperature $T$ at constant volume $V$:
\be \label{specheat}
C_V \, = \, \left(\frac{\partial E}{\partial T}\right)_{V} \, .
\ee
The gravitational waves, in turn, are given by 
\be \label{gravwaves}
{C^{ij}}_{ij} \, = \, \bigr< {T^i_j}^2 \bigl> - \bigr< T^i_j \bigl>^2 \,\,\, i \neq j
\ee
where the indices $i$ and $j$ are related to the polarization tensor of the wave which is being considered.

In the case of String Gas Cosmology, the thermal correlation functions for a gas of closed strings in the high temperature Hagedorn phase have holographic scaling, i.e. $C_V \sim R^2$, and correspondingly for the other correlation functions \cite{Deo}. This result can be understood heuristically from the fact that strings look like point particles in one lower spatial dimension. This then leads to the scale-invariance of the spectrum of cosmological perturbations and gravitational waves.  The fact that the temperature is a slightly decreasing function of time towards the end of the Hagedorn phase (when scales exit the Hubble radius) leads to a slight red tilt of the spectrum of cosmological perturbations (modes with larger values of $k$ exit the Hubble radius later). The fact that the pressure is an increasing function of time towards the end of the Hagedorn phase leads to a characteristic slight blue tilt in the spectrum of gravitational waves \cite{NBPV} \footnote{Recall that inflation models in the context of General Relativity (with matter obeying the usual energy conditions) always leads to a slight red tilt of the spectrum of gravitational waves. This characteristic blue tilt of the spectrum of gravitational waves is also obtained \cite{Ziwei} in the recently proposed version of the Ekpyrotic scenario in which an S-brane motivated by superstring theory leads to a nonsingular transition between an Ekpyrotic contracting phase and the radiation phase of the SBB.}.

\subsection{Cosmological Perturbations in Matrix Cosmology}

We will now perform the calculation of the spectra of cosmological perturbations and gravitational waves in our scenario. Specifically, we are interested whether a scale-invariant spectrum of cosmological perturbations emerges \footnote{In the case of the proposal of \cite{Vafa} such a spectrum emerges because of the conformal invariance of the emergent topological phase.}. We cannot apply the abovementioned heuristic argument for such a spectrum since our calculation is not based on classical string degrees of freedom. On the other hand, since our scenario in the perturative limit will reduce to perturbative string theory, it would not be surprising if the holographic scaling of the specific heat capacity emerges.

Our starting point is the finite temperature action $S(\beta)$ (\ref{BFSSbosonic}) of the BFSS matrix theory. The resulting finite temperature partition function is given by the functional integral
 \begin{eqnarray}
 Z(\beta) \, = \, \int \[\mathcal{D}A\]_\beta \[\mathcal{D}X\]_\beta\; e^{-S(\beta)} \,,
\end{eqnarray}
where the subscript $\beta$ implies that the fields to be integrated are over this range. Given this partition function, the internal energy $E$ is given by
\begin{eqnarray}
E \, = \, -\frac{\d}{\d\beta} \ln Z(\beta)\,.
\end{eqnarray}
One can now calculate this internal energy as follows:
\begin{eqnarray}
	E \, = \, -\frac{1}{Z(\beta)} \lim_{\Delta\beta\rightarrow 0} \frac{Z(\beta') - Z(\beta)}{\Delta\beta}\,,
\end{eqnarray}
where $\beta'=\beta +\Delta\beta$. $Z(\beta')$ is given by
\begin{eqnarray}
	Z(\beta') = \int \[\mathcal{D}A'\]_\beta' \[\mathcal{D}X'\]_\beta'\; e^{-S(\beta')} \,,
\end{eqnarray}
where $S'$ can be obtained from the action by replacing $\beta,\, t,\, A(t),\, X_i(t)$ with $\beta',\, t',\, A'(t'),\, X'_i(t')$.  While the measures remain invariant, $\[\mathcal{D}A'\]_\beta' = \[\mathcal{D}A\]_\beta$ and $\[\mathcal{D}X'\]_\beta' = \[\mathcal{D}X\]_\beta$, the fields and time-parameter are related by the transformations
\begin{eqnarray}
	t' = \frac{\beta'}{\beta} t \,, \hspace{2mm} A'(t') = \frac{\beta}{\beta'} A(t)\,, \hspace{2mm} X_i'(t') = \sqrt{\frac{\beta'}{\beta}} X_i(t)\,.
\end{eqnarray}	
Under this transformation, the kinetic term remains invariant while the interaction term in $S'$ is related to that in $S$ as follows:
\begin{widetext}
\begin{eqnarray}
	\int_0^{\beta'} \d t' \, {\rm Tr}\left\{\[X'_i(t'), X'_j(t')\]^2\right\} = \(\frac{\beta'}{\beta}\)^3 \;	\int_0^\beta \d t \, {\rm Tr}\left\{\[X_i(t), X_j(t)\]^2\right\} \,.
\end{eqnarray}
One can now write down the relation
\begin{eqnarray}
	Z(\beta') &=& \int \[\mathcal{D}A'\]_\beta' \[\mathcal{D}X'\]_\beta'\; e^{-S} \; \exp\( -N\int_0^\beta \d t \; \frac{1}{4} {\rm Tr} \left\{ \[X_i, X_j\]^2 -  \(\frac{\beta'}{\beta}\)^3  \[X_i, X_j\]^2\right\} \) \nn\\
	&=& \int \[\mathcal{D}A'\]_\beta' \[\mathcal{D}X'\]_\beta'\; e^{-S} \left\{ 1 - \frac{3}{4} N \int_0^\beta \d t \; {\rm Tr} \( \[X_i, X_j\]^2\) \frac{\Delta\beta}{\beta} + \mathcal{O}(\(\Delta\beta\)^2 \right\}\nn\\
	&=& Z(\beta) \(1 - N^2 \Delta\beta \langle \mathcal{E} \rangle + \mathcal{O}(\(\Delta\beta\)^2)\)\,,
\end{eqnarray}
\end{widetext}
where 
\begin{eqnarray}
	\mathcal{E} \, = \, -\frac{3}{4} \frac{1}{N\beta} \int_{0}^{\beta} \d t \; {\rm Tr}\(\[X_i, X_j\]^2\)\,.
\end{eqnarray}
 Therefore, we can write down the internal energy as
\begin{eqnarray}
	E \, = \, N^2 \langle\mathcal{E}\rangle\,,
\end{eqnarray}
where $\langle\cdot\rangle$ is the expectation value calculated with respect to the partition function $Z(\beta)$.

It has been shown in \cite{equiv} that the BFSS action reduces to the IKKT one at high temperatures. If one Fourier expands the fields as
\begin{eqnarray}
	X_i \, = \, \sum_n X_i^n \, e^{in\omega t}\,,
\end{eqnarray}
where $\omega = 2\pi/\beta$ are the Matsubara frequencies, then the BFSS action becomes
\begin{eqnarray}
	S_{\rm BFSS} \, = \, S_0 +S_{\rm kin} + S_{\rm int}\,.
\end{eqnarray}
Let us first consider the leading order behaviour of the action, in the high temperature limit, which is given by the zero modes of the Fourier expansion, and therefore 
\begin{eqnarray}
	S_0 \, \equiv \, -N\beta \; {\rm Tr}\left\{\frac{1}{2} \(\[A, X_i^0 \]\)^2 +\frac{1}{4}\(\[X_i^0, X_j^0\]\)^2 \right\}\,.	
\end{eqnarray}
We rescale the zero modes as \footnote{Note the consistency of the mass dimensions using the conventions introduced at the beginning of Section II.}
\begin{eqnarray}
	A_i := T^{-1/4} X_i^0 \; (i=1,2,\ldots,d), \hspace{2mm} A_D := T^{-1/4} A\,,
\end{eqnarray}
where $D = d+1$. This tells us that
\begin{eqnarray}
	S_0 = \frac{1}{4} \, N\; {\rm Tr}\(F_{\mu\nu}\) =: S_{\rm IKKT}\,, \hspace{2mm} F_{\mu\nu} := -i \[A_\mu,A_\nu\]\,.
\end{eqnarray}
Here we assume $(\mu,\nu =1,2,\ldots, D)$. On the other hand, the kinetic term of the action go as
\begin{eqnarray}\label{S_kin}
S_{\rm kin}
&\equiv&
N \beta \ {\rm tr} 
\bigg\{ \frac{1}{2} \sum_{n \not=0} (n\omega)^2
X^i_{-n} X^i_{n}\bigg\}
\end{eqnarray} 
and and the interaction terms as
\begin{widetext}
\begin{eqnarray}\label{S_int}
	S_{\rm int} = -	N \beta \ {\rm tr} 	\bigg\{	\sum_{n \neq 0}	n \omega X^i_{-n} [A,X^i_{n}]
	+\frac{1}{2} \sum_{n \not= 0} [A,X^i_{-n}][A,X^i_{n}] +\frac{1}{4}	\sum_{npq} \,
	[X^i_{-n-p-q},X^j_n][X^i_p,X^j_q]
	\bigg\} \,, 
\end{eqnarray}
\end{widetext}
where the $n=p=q=0$ term is excluded in the last sum. We only show the bosonic terms above and ignore the terms corresponding to the fermionic and the ghost fields to avoid clutter (see \cite{equiv} for more details). While we use $S_0$ to calculate the leading order terms, $S_{\rm kin}$ and $S_{\rm int}$ become important when calculating the next-to-leading order terms.

Given this, we can calculate two quantities of interest to us -- the extent of the eigenvalue distribution and the internal energy. The extent of the eigenvalue distribution is given by  
\begin{eqnarray}\label{Spatial_extent}
	R^2 := \frac{1}{N\beta} \int_0^\beta \d t \; {\rm Tr}\(X_i(t)\)^2 \, ,
\end{eqnarray}
We begin by taking its expectation value with the BFSS partition function and we find that
\begin{eqnarray}\label{range}
	\langle R^2 \rangle_{\rm BFSS} \simeq \chi_1 T^{1/2}\,,
\end{eqnarray} 
where 
\begin{eqnarray}
	\chi_1 := \left\langle \frac{1}{N} \; {\rm Tr}(A_i)^2\right\rangle_{\rm IKKT}\,.
\end{eqnarray}
Note that although expression \eqref{Spatial_extent} is an exact definition, \eqref{range} is the leading order (in temperature) relation between $R$ and $\chi_1$. The leading term is given by the zero modes. Note that the extent of space parameter in the IKKT model is given by the quantity $\chi_1$ defined above (compare with \eqref{IKKT_R} above). The dependence of the extent of eigenvalue distribution on the spatial volume is characterized by $\chi_1$, whereas we explicitly separate its dependence on temperature as above. The numerical simulations of the IKKT model tell us how the extent of space parameter \eqref{IKKT_R} evolves with time. The temperature (coming from the Euclideanized time direction) can be assumed to be constant in the model and $\chi_1$ will give us the value of the background volume for this given time.


More relevantly for us, the internal energy $E =   N^2 \langle\mathcal{E}\rangle_{\rm BFSS}$ can be calculated to get
\begin{eqnarray}\label{E1}
	E \simeq \frac{3 N^2}{4}\, \chi_2\, T\,,
\end{eqnarray}
 where
\begin{eqnarray}
	\chi_2 := \left\langle\frac{1}{N} \, {\rm Tr}(F_{ij})^2\right\rangle_{\rm IKKT}\,.
\end{eqnarray} 
This shows that the internal energy, to the leading order in temperature, does not depend on the spatial volume (as this expression does not depend on $\chi_1$).

Let us make a quick note regarding the dimensions involved in the above equations. Recall that since we have set the 't hooft coupling $\lambda = 1$, we need to replace factors of $N$ by $1/g^2$, which has mass dimension of $-3$. The Yang-Mills coupling $g$ is related to the string length $l_s$ by the dimensionless string coupling constant $g_s$, via $g^2 = g_s/l_s^3$. Therefore, for restoring appropriate dimensions, one needs to insert proper powers of $g$ in the expressions above. For instance, for the internal energy to have mass dimension $1$, we therefore require that $\chi_2 \sim l_s^{-6}$ (due to the factor of $N^2 \sim 1/g^4$ in \eqref{E1} above). Similarly, the extent of space parameter $R^2$ must have a factor of $l_s^{7}$ in its definition \eqref{range} for it to have mass dimension of $-2$. We continue to suppress these dimensional parameters for simplicity and would reinstate them in the final expression for the power spectra.

Going to the next to leading order in temperature, one can calculate the internal energy to be (we follow the conventions of \cite{equiv}):
\begin{eqnarray}\label{E}
	E &\simeq&  \frac{3 N^2}{4}\, \chi_2\, T \\
	&& - \frac{3 N^2}{4} \left(\frac{d-1}{12} - \frac{p}{8}\right)\,\left(\chi_5 -\chi_6 - 4\chi_1\right)\, T^{-1/2}\,,\nonumber
\end{eqnarray}
where 
\bea
\chi_5\, &:=&  \left\langle{\rm Tr}\left(F_{ij}\right)^2 .\, {\rm Tr}\left(A_{k}\right)^2 \right\rangle_{\rm IKKT} \nonumber \\
\chi_6 \, &:=&  \left\langle{\rm Tr}\left(F_{ij}\right)^2 .\, {\rm Tr}\left(A_{D}\right)^2 \right\rangle_{\rm IKKT}\,.
\eea
In the above expression, $p$ denotes the number of fermionic superpartners of the $d$ bosonic matrices. 

Let us calculate $\chi_5$ and $\chi_6$ using the following approximation
\bea
 \chi_5 \, &=&  \left\langle{\rm Tr}\left(F_{ij}\right)^2 .\, {\rm Tr}\left(A_{k}\right)^2 \right\rangle_{\rm IKKT} \nonumber \\
&\simeq&  \left\langle{\rm Tr}\left(F_{ij}\right)^2\right\rangle_{\rm IKKT} \, \left\langle{\rm Tr}\left(A_{k}\right)^2 \right\rangle_{\rm IKKT} \nonumber \\
  &=& \, N^2 \,\chi_2 \,\chi_1
\eea
and
\bea  
 \chi_6 \, &=&  \left\langle{\rm Tr}\left(F_{ij}\right)^2 .\, {\rm Tr}\left(A_{D}\right)^2 \right\rangle_{\rm IKKT} \nonumber \\
 &\simeq& \left\langle{\rm Tr}\left(F_{ij}\right)^2\right\rangle_{\rm IKKT}\, \left\langle{\rm Tr}\left(A_{D}\right)^2\right\rangle_{\rm IKKT} \nonumber \\
 &=& \, \left(\frac{N^2}{d}\right)\, \chi_2  \, \chi_1 \, .
\eea
 
Let us make two observations regarding the above calculation. Firstly, the next to leading order values for these quantities can be evaluated explicitly by considering the propagators, from the kinetic term in the action, and from the interaction terms. However, although we only showed the bosonic terms in \eqref{S_kin} and \eqref{S_int} for simplicity, one also needs to take into account the fermionic fields (and the ghost terms corresponding to our gauge-fixing) to carry out the explicit calculation. And finally, one needs to integrate out \textit{only} over the non-zero modes in order to arrive at the above-mentioned results. The zero modes (in the Matsubara frequencies) are what gives rise to the IKKT action and therefore, we express our results in terms of quantities evaluated in the IKKT model and the temperature $T$. Note that since we have set the 'tHooft coupling $\lambda = g^2 N = 1$, our only dimensionful parameter for perturbation theory is $T^{-3/2}$ \cite{equiv2}. In other words, once one integrates out non-zero modes using perturbation theory, the leftover integration over the zero modes can be thought of as taking the expectation value of Green's functions using the bosonic part of the IKKT action.  

We now have the ingredients needed to evaluate the power spectrum of energy density fluctuations in our scenario using (\ref{eflucts}) and (\ref{specheat}).  Let us consider a comoving momentum scale $k$. The associated volume is $\langle R^2 \rangle_{\rm BFSS}^{3/2}$ which we will in the following abbreviate by $R^3$. The dimensionless power spectrum $P(k)$ on the scale $R$ related to the wavenumber $k$ via $R = 2 \pi k^{-1}$ is given by
\begin{eqnarray}
	P(k) \, &\sim&  \, k^3 \langle\vert \Phi(k) \vert^2\rangle \nonumber \\
	&=& \, 16 \pi^2 G^2 k^{-1} \langle \delta T^0_0(k) \delta T^0_0(k) \rangle \\
	&=& \, 16 \pi^2 G^2 k^{-4} ( \delta \rho )^2 \nonumber \\
	&=& \, 16 \pi^2 G^2 k^{-4} T^2 C_V R^{-6} \, , \nonumber \\
	&=& \, 16 \pi^2 G^2 k^2 T^2 C_V (kR)^{-6} \nonumber
\end{eqnarray} 
where the factor of $k^{-3}$ in going from the second to the third line comes from converting momentum space to position space density. 
 
Thus, the scalar power spectrum depends mostly on the specific heat $C_V$. Let us calculate it to the next-to-leading order in the high temperature limit. From \eqref{E}, we find
\begin{eqnarray}
	C_V \, &=& \,  \frac{3 N^2}{4} \chi_2  \\
	&+& \, \dfrac{3 N^4}{8} \left(\frac{d-1}{12}-\frac{p}{8}\right) \,\left(\chi_2 - \frac{1}{d}\chi_2 - \frac{4}{N^2}\right) \, \chi_1\, T^{-3/2}\, . \nonumber
\end{eqnarray}
Note that the $C_V > 0$ for $d=3,\, p=4$ and the thermodynamics is well-defined in this case. 

The first term hence yields a contribution to the power spectrum proportional to $k^2$. Since $\chi_1 \sim k^{-2}$, the second term yields a scale invariant contribution.  On microscopic scales, the first term dominates. It corresponds to a Poisson spectrum and is what we expect for thermal fluctuations on scales close to the correlation length of the system. On infrared scales relevant for current cosmological fluctuations, however, it is the second term which dominates, and it corresponds to a scale-invariant spectrum, and its value is
\bea \label{scalar}
P(k) \, &=& \, 16 \pi^2 G^2 k^2 (kR)^{-6} T^{1/2} N^2 \chi_1 \frac{3}{8} \\
& & \bigr(\frac{d -1}{12} - \frac{p}{8} \bigl) \bigr(N^2 \chi_2 - \frac{N^2}{d} \chi_2 - 4 \bigl) \, .
\nonumber
\eea
Substituting for $\chi_1$ making use of (\ref{range}), and reinstating dimensional parameters, yields
\bea \label{scalar2}
P(k) \, &=& \, 16 \pi^2\,(kR)^{-4} \left(\frac{1}{l_s m_{pl}}\right)^4 \left(\frac{3}{8}\right) \\
& & \left(\frac{d -1}{12} - \frac{p}{8} \right) \left(\frac{(d-1)^2}{d}\left(1 - \frac{1}{N^2}\right) - 4 \right) \, ,
\nonumber
\eea
from which it follows that the amplitude of the spectrum is given by
\be
{\cal{A}} \, \sim \, (l_s m_{pl})^{-4} \, ,
\ee
the same scaling as in String Gas Cosmology \cite{NBV}. In \eqref{scalar2}, we have used the explicit expression for $\chi_2$ \cite{chi2}:
\begin{eqnarray}\label{chi2}
	\chi_2 \,= \, (d-1) \(1 - \frac{1}{N^2}\)\,,
\end{eqnarray}
where all dimensional factors have been accounted for and there is no further $g$ dependence coming from the $N^2$ term. Note that this result does not depend on the exact dynamics of how the background volume expands with time (beyond the general evidence the numerical analysis provides for the emergence of $3$ large spatial dimensions).

\subsection{Gravitational Waves in Matrix Cosmology}

Tensor perturbations are sourced by the off-diagonal pressure perturbations, as described in (\ref{gravwaves}). Specifically, the dimensionless power spectrum of gravitational waves on a comoving momentum scale $k$ is given by
\be
P_h(k) \, = \, 16 \pi^2 G^2 k^{-4} C^{ij}_{ij}(R(k)) \, ,
\ee
where we recall that $C^{ij}_{ij}(R(k))$ is the position space expectation value of the square of the off-diagonal pressure perturbation ($i \neq j$), and $R(k)$ is the length scale corresponding to $k$. In a thermal state we expect the off-diagonal pressure perturbations to be smaller but of similar magnitude as the diagonal pressure contribution. We will denote the suppression factor of the off-diagonal term compared to the diagonal term by a positive constant $\alpha < 1$. Hence,
\be
C^{ij}{}_{ij} \, = \, \alpha \frac{T}{R^2} \frac{\partial \tilde{p}}{\partial R} \, ,
\ee
where the pressure $\tilde{p}$ is given by
\be
\tilde{p} \, = \, - \frac{1}{V} \frac{\partial {\cal{F}}}{\partial {\rm{ln}} R} \, .
\ee

To calculate the pressure, let us begin with the free energy of our system, calculated up to next-to-leading order
\bea
{\cal{F}} \, &=& \,  \frac{3N^2}{4\beta} \left[ \chi_2 \,{\rm{ln} \beta} \right.\\
& &\left. \, -\frac{2}{3}\, \left( \frac{d - 1}{12} - \frac{p}{8} \right)\left(\chi_5 - \chi_6 - 4\chi_1\right)  \beta^{3/2} \right] \, . \nonumber
\eea
We can use the same approximations as before to write $\chi_5$ and $\chi_6$ in terms of $\chi_1$ and $\chi_2$. We then obtain
\be
C^{ij}{}_{ij} \, = \, \alpha  \frac{T^{1/2}}{R^4} N^2 \bigr( \frac{d - 1}{12} - \frac{p}{8} \bigl)(N^2 \chi_2 - \frac{N^2}{d} \chi_2 - 4) \, ,
\ee
from which it follows that the  dimensionless power spectrum of gravitational waves will also be scale-invariant with an amplitude $P_h(k)$ given by (on restoring the dimensional factors, and using \eqref{chi2}, as before):
\bea \label{tensor}
P_h(k) \, &=& \, \alpha\, 16 \pi^2\,(kR)^{-4} \left(\frac{1}{l_s m_{pl}}\right)^4 \left(\frac{3}{8}\right) \\
& & \left(\frac{d -1}{12} - \frac{p}{8} \right) \left(\frac{(d-1)^2}{d}\left(1 - \frac{1}{N^2}\right) - 4 \right) \, .
\nonumber
\eea

Comparing the results (\ref{tensor}) and (\ref{scalar}) for the tensor and scalar power spectra, we find that the tensor to scalar ratio $r$ is given by
\be
r \, = \,  \frac{8}{3} \alpha \, .
\ee
In order to be consistent with the current observational bound on $r$, the value of $\alpha$ needs to be of the order $\mathcal{O}\left(10^{-2}\right)$ or smaller.  Note that although the off-diagonal elements are naturally suppressed compared to the diagonal ones, for thermal fluctuations, they are not expected to get fine-tuned to be extremely small. In other words, we expect the parameter $\alpha$ to be a smaller than $1$ but not by many orders of magnitude \cite{NBV}. Note that in String Gas Cosmology the value of $r$ is suppressed by the ratio between the pressure and the energy density in the Hagedorn phase \cite{NBPV}. In the topological phase model of \cite{Vafa}, no primordial gravitational waves are generated to leading order in the analysis. However, since $\alpha$ is not expected to be many orders of magnitude smaller than $1$ for our model, we expect to find an observable signal for primordial gravity waves in our model. This is a significant difference between our model and those other approaches to early universe cosmology. It is hence important to estimate the value of $\alpha$ which results from our matrix theory model.

\section{Conclusions and Discussion}

In this paper we have suggested a concrete realization of a non-singular {\it emergent cosmology} based a matrix theory, a proposed non-perturbative definition of superstring theory in which space is emergent. The starting point is a gauge action for nine Hermitean $N \times N$ matrices $X_i$. The covariant derivative involves another $N \times N$ matrix $A$.  We consider this matrix model in a finite temperature state. Space is emergent in the sense that in the large $N$ limit, the expectation values of $X_i^2$ yield the size of the i'th spatial dimension. We have used results of numerical and analytical studies of matrix theory to show that a spontaneous breaking of the $SO(9)$ spatial symmetry takes place, and that exactly three dimensions of space become large. We have argued that at late times, a phase transition to the radiation phase of Standard Big Bang cosmology takes place, signalling the end of the emergent phase. Our scenario automatically solves some problems of Standard Big Bang cosmology such as the horizon problem, in the same way that they are solved in the proposal of \cite{Vafa}. A quick way to see this is to realize that the emergent spatial dimensions appear from the early non-geometric phase when the matrices are not commuting and their eigenvalues cannot be said to describe a smooth $(3+1)$-d spacetime. Thus, the entire emergent space is born out of the same matrix action and is interacting with each other in the non-geometric phase, naturally resolving the horizon problem \footnote{Since space is emergent, the very concept of causality is also emergent in this theory.}.

However, background dynamics is not sufficient for understanding the properties of the emergent cosmology derived from matrix theory. One needs to calculate the spectrum of primordial perturbations in this model and this is where the novelty of our work lies. We have computed the thermal correlation functions of the energy-momentum tensor in the emergent phase. These determine the spectrum of cosmological density fluctuations and gravitational waves. In analogy to what is assumed in String Gas Cosmology, the fluctuations are of thermal origin. They do not originate as quantum vacuum perturbations as they do in canonical inflationary models. We find that the spectrum of cosmological fluctuations have two components, one of which has Poisson scaling and dominates on small scales, the other one being scale-invariant which dominates on scales relevant to cosmological observations. The spectrum of gravitational waves is also scale-invariant. We have computed the tensor to scalar ratio $r$ on large scales. The resulting amplitude is given by the ratio of the off-diagonal to the diagonal pressure fluctuations, a ratio which we denote by $\alpha$ in the text. In order not to exceed the observational upper bound on $r$, the value of $\alpha$ needs to be sufficiently small. An open problem is to derive the value of $\alpha$ from our matrix theory model. 

Note that the spectrum of both density perturbations and primordial tensor modes is not expected to be exactly scale-invariant on observable scales. Small deviations from scale-invariance, and a corresponding small tilt, naturally appear in our model when one goes to the next order in temperature. In addition, the processing of the fluctuations through the phase transition can induce a tilt, as it does in String Gas Cosmology. One can calculate the bispectrum and other higher order moments from higher order calculations in perturbation theory for the thermal state under consideration. We leave these topics for future work. 

Note that our scenario does not involve a period of inflationary expansion. Since it is based on a non-perturbative approach to superstring theory, the scenario is free from any {\it swampland} constraints, consistency conditions which rule out many inflationary models (see \cite{swamp1} for reviews of the swampland program, and \cite{swamp2} for applications to inflation). The scenario is clearly consistent with the {\it trans-Planckian censorship conjecture} (TCC) \cite{Bedroya1} since the wavelengths of fluctuation modes which we observed today were never smaller than the Planck length (the fluctuations are generated towards the end of the emergent phase on scales which are macroscopic compared to the string length). This is another difference compared to the inflationary scenario, where the TCC sets a very restrictive upper bound on the energy scale of inflation \cite{Bedroya2}, a bound which most models of inflation fail to satisfy.

Lastly, note that there are other approaches to obtaining space-time and cosmology from matrix theory. For example, Steinacker has a research program (see \cite{Stein1} for some original articles and \cite{Stein2} for a review) in which matrices satisfying the equations of motion derived from the matrix action are represented on a Poisson manifold. Specifically, one can choose the Poisson manifold to have space-time dimension four. This corresponds to choosing a background matrix set with $X^{a} = 0$ for $a \neq 0, 1, 2, 3$. Matrix fluctuations about this background then yield an action for gauge fields and scalar fields, and fermions if one starts from a supersymmetric matrix model. Gravity is induced on the background. Our work is different in that we obtain space-time directly from the matrix theory.

We would also like to mention recent work of Klinkhamer \cite{Klinkhamer} which further develops some of the ideas of \cite{IKKT2} for the Lorentzian matrix model, argues that matrix theory will yield a nonsingular emergent cosmology, and extracts space and the cosmological scale factor from a numerical analysis of the model. However, no attempt is made to compute cosmological perturbations and compare with observations.

The most important open issue for our scenario is the study of the transition from the emergent phase analyzed in this paper to the radiation phase of Standard Big Bang cosmology. In the case of String Gas Cosmology, the transition proceeds via the annihilation of string winding modes, resulting in the generation of string loops which lead to radiation. The transition is smooth and a high density radiation bath is automatically generated, obviating the need of a separate reheating phase, a phase which is needed in inflationary cosmology. In the same way, in our scenario the exit from the emergent phase will automatically lead to a high density radiation bath. The details of the transition, however, are not known, and these details will be important in order to be able to make precise predictions for the slopes of the spectra of scalar and tensor modes. Work on this issue is in progress.

\section*{Acknowledgments}

The research at McGill is supported in part by funds from NSERC and from the Canada Research Chair program. SB is supported in part by the NSERC (funding reference CITA \#490888-16) through a CITA National Fellowship and by a McGill Space Institute fellowship SL is supported in part by funds from a Templeton Foundation sub-contract. We are grateful to Sumit Das and Keshav Dasgupta for comments on a draft of this paper.

\end{document}